\documentclass[aps,prc,twocolumn,superscriptaddress,showpacs,showkeys]{revtex4}
\usepackage{amsmath,amssymb}
\usepackage{bm}
\usepackage[dvips]{graphicx}
\newcommand{\Tr}{\text{Tr}}

\begin{document}


\title{Flavor SU(3) breaking effects in the chiral unitary model
for meson-baryon scatterings}

\author{T.~Hyodo}%
\affiliation{Research Center for Nuclear Physics (RCNP), 
Ibaraki, Osaka 567-0047, Japan}%
\author{S.~I.~Nam}%
\affiliation{Research Center for Nuclear Physics (RCNP), 
Ibaraki, Osaka 567-0047, Japan}%
\affiliation{Department of Physics,
Pusan National University, Pusan 609-735, Korea}%
\author{D.~Jido}%
\affiliation{Research Center for Nuclear Physics (RCNP), 
Ibaraki, Osaka 567-0047, Japan}%
\author{A.~Hosaka}%
\affiliation{Research Center for Nuclear Physics (RCNP), 
Ibaraki, Osaka 567-0047, Japan}%

\date{\today}

\begin{abstract}
    We examine flavor SU(3) breaking effects on meson-baryon
    scattering amplitudes in the chiral unitary model.  
    It turns out that the SU(3) breaking,
    which appears in the leading quark mass term in 
    the chiral expansion, can not explain the channel dependence of
    the subtraction parameters of the model, which are
    crucial to reproduce the observed scattering amplitudes
    and resonance properties.  
\end{abstract}

\pacs{12.39.Fe, 11.80.Gw, 14.20.Gk, 14.20.Jn, 11.30.Hv}
\keywords{chiral unitary approach, meson-baryon scatterings,
flavor SU(3) breaking}

\maketitle

Properties of baryonic excited states
are investigated with great interest 
both theoretically and experimentally.  
Recently, the chiral unitary model has been successfully applied to
this problem, especially to the first excited states of negative
parity ($J^P = 1/2^-$) such as $\Lambda(1405)$ and
$N(1535)$~\cite{Kaiser:1995cy,Krippa:1998us,
Oset:1998it,Lutz:2001yb,
Oset:2001cn,Inoue:2001ip,Ramos:2002xh}.  
In this method, based on the leading order interactions of the 
chiral Lagrangian and the unitarization of the S-matrix,
the baryon resonances are dynamically generated
as quasi-bound states of 
ground state mesons and baryons.  
It reveals the importance of chiral dynamics not only in the
threshold but also in the resonance energy region.  

In the chiral unitary model for the meson-baryon scattering, 
we consider the coupled channel scatterings of the octet mesons
and baryons. Imposing the unitarity condition on the
scattering amplitudes $T_{ij}$ in the $N/D$ method, we obtain
the scattering equation in the matrix form~Refs.~\cite{Oset:1998it,Oller:2000fj}:
\begin{equation}
    T_{ij} = V_{ij} + V_{ik} G_kT_{kj} \ , 
    \label{eq:BS}
\end{equation}
where $V_{ij}$ denotes
the elementary tree level interaction
derived from the chiral Lagrangian.
This equation can be 
solved algebraically.
The loop integral $G_{i}$ is the fundamental building block in the 
chiral unitary model and are regularized by the dimensional regularization;
\begin{align}
    G_i(\sqrt{s})=& i\int\frac{d^{4}q}{(2\pi)^{4}}
    \frac{2M_{i}}{(P-q)^{2}-M_{i}^{2}+i\epsilon} 
    \frac{1}{q^{2}-m_{i}^{2}+i\epsilon} \nonumber \\ 
    =&\frac{2M_{i}}{(4\pi)^{2}}
    \Bigl[a_i(\mu)+\ln\frac{M_i^{2}}{\mu^{2}}
    +\frac{m_i^{2}-M_i^{2}+s}{2s}\ln\frac{m_i^{2}}{M^{2}_i} \nonumber \\
    &+\frac{\bar{q}_i}{\sqrt{s}}
    ({\rm Ln}_{+-}+{\rm Ln}_{++}-{\rm Ln}_{-+}-{\rm Ln}_{--}
    )\Bigr],
     \label{eq:loop}
\end{align}
with ${\rm Ln}_{\pm\pm}\equiv \ln(\pm s\pm 
(M^{2}_i-m_i^{2})+2\sqrt{s}\bar{q}_i) )$,
the masses of baryon and meson $M_i$ and $m_i$, the three-momentum
of the meson $\bar{q}_i$, the total energy in the center of mass system
$\sqrt{s}$
and the regularization scale $\mu$.
In the present calculation, we follow the 
method shown in refs.~\cite{Oset:2001cn,Inoue:2001ip,Hyodo},
and calculate only the $s$-wave amplitudes since the contributions 
from the $p$-wave (and higher partial waves) are much less important
in energies considered here~\cite{Jido:2002zk}.

In actual calculations, it is necessary to determine the renomalization
constants ($a_{i}$'s) in Eq.\eqref{eq:loop} so as to reproduce
experimental data. The constants $a_{i}$ are equivalent to the
subtraction constants in the dispersion theory formulation~\cite{Oller:2000fj}
and,  in fact, are free parameters of the model. 
As a consequence,
they have depended very strongly 
on scattering channels, as shown in Table~\ref{tbl:subtractions}.

\begin{table}
    \centering
\begin{ruledtabular}
    \begin{tabular}{ccccccc}
    $S=-1$ & $\bar{K}N$ & $\pi\Sigma$ & $\pi\Lambda$
    & $\eta\Lambda$ & $\eta\Sigma$ & $K\Xi$  \\
    \hline
    $a_i$& $-1.84$ & $-2.00$ & $-1.83$
    & $-2.25$ & $-2.38$ & $-2.67$  \\
    \end{tabular}
   \vspace{0.1cm}
   \begin{tabular}{ccccc}
    $S=0$ & $\pi N$ & $\eta N$ & $K\Lambda$ & $K\Sigma$  \\
    \hline
    $a_i$ & 0.711 & $-1.09$ & 0.311 & $-4.09$  \\
    \end{tabular}
\end{ruledtabular}
    \caption{Channel dependent subtraction constants  $a_i$ obtained in
    Refs.~\cite{Oset:2001cn,Inoue:2001ip}
    with $\mu=630$ MeV. 
    }
    \label{tbl:subtractions}
\end{table}%

In this work, we investigate whether such
channel dependence of the subtraction constants could
be explained by the SU(3) breaking terms 
of the chiral perturbation theory.
By doing this,
we expect that the free parameters of $a_i$'s
could be controlled with suitable physics ground,
namely the SU(3) breaking terms,
in order to extend this model to various channels 
with predictive power.
Here we keep using just one subtraction constant $a$ 
commonly in all channels
to regularize the loop function $G_i$.

The use of only one subtraction constant
is justified in the SU(3) symmetric limit~\cite{Hyodo,Jido:2003cb}.
Under the flavor SU(3) symmetry,
the scattering amplitude should be expressed 
as a diagonal matrix in the SU(3) basis ($1,8,\cdots$),
which is transformed from the particle basis ($\pi N, \eta N, \cdots$)
by a fixed unitary matrix given by the SU(3) Clebsch-Gordan coefficients.
Each components of the amplitude $T(D)$ separately 
satisfies the scattering equation like Eq.\eqref{eq:loop} in each irreducible 
representation $D$.
Therefore, on one hand, 
the function $G$ represented in a matrix form
becomes a diagonal matrix in the SU(3) basis.
On the other hand, since the $G$ function is given as a 
loop integral as shown in Eq.~\eqref{eq:loop}, 
it is also diagonal in the particle basis.  
Therefore the subtraction constants $a_i$'s
are components of a diagonal matrix both in the SU(3) and particle bases.
Such a matrix for the subtraction constants
should be proportional to unity.  
Hence, it is concluded that 
there is only one subtraction constant 
$a$ in the SU(3) limit.

Now let us show the Lagrangian with the flavor SU(3) breaking terms,
which we use in the present work.
The SU(3) breaking appears as the quark mass terms 
in the chiral expansion:
\begin{align}
     \begin{split}
	\mathcal{L}_{SB}
	=&-\frac{Z_0}{2}\Tr \Bigl(
	d_m\bar{B}\{\xi\mathbf{m}\xi
	+\xi^{\dag}\mathbf{m}\xi^{\dag},B\} \\
	&+f_m\bar{B}[\xi\mathbf{m}\xi
	+\xi^{\dag}\mathbf{m}\xi^{\dag},B]
	\Bigr) \\
	&-\frac{Z_1}{2}\Tr(\bar{B}B)
	\Tr(\mathbf{m}U+U^{\dag}\mathbf{m}) \ .
     \end{split}
     \label{eq:SBLag}
\end{align}
where $f_{m}+d_{m}=1$.
Here we employ the standard notation~\cite{Donoghue:1992dd}:
$\xi(\Phi)  =\exp\{i\Phi/\sqrt{2}f\}$ and
$U(\Phi) = \xi^2
$. The $3 \times 3$ matrices $B$ and $\Phi$ represent
the baryon and meson fields. 
At this stage, we introduce one meson decay constant $f$,
which is taken as an averaged value $f=1.15f_{\pi}$ with $f_{\pi}=93$ MeV.
The quark mass matrix is defined as
$\mathbf{m}=\text{diag}(\hat{m},\hat{m},m_s)$ with
isospin symmetry, $m_u=m_d\equiv \hat{m}$. 
The parameters 
$Z_0,Z_1, f_m/d_m$ can be determined by the baryon masses 
and the $\pi N$ sigma term, and therefore we have no new free parameters. 
Here we have
$Z_0 = 0.528$, $Z_1=1.56$ and $f_m / d_m =-0.31$
with $m_s/\hat{m}=26$,
which are determined in chiral perturbation theory for meson masses.
The terms in Eq.~\eqref{eq:SBLag} are of order ${\cal O}(p^2)$,
based on the 
Gell-Mann-Oakes-Renner relation~\cite{Gell-Mann:1968rz},
which implies $m_q \propto m_{\pi}^2$.  
There are other chirally symmetric 
terms of order ${\cal O} (p^2)$.  
Here we do not take into account these terms, since 
we concentrate on the effects of the flavor SU(3) breaking.

Let us show the numerical results of the 
$\bar{K}N$ induced scatterings.  
We use a single subtraction constant $a$,
and compare the results with and without the SU(3) breaking terms.
In each case, the subtraction constant is determined 
by fitting threshold branching ratios~\cite{Nowak:1978au,Tovee:1971ga}:
\begin{equation}
    \begin{split}
	\gamma=&\frac{\Gamma(K^{-}p\to\pi^{+}\Sigma^{-})}
	{\Gamma(K^{-}p\to\pi^{-}\Sigma^{+})}\sim  2.36\pm0.04 \ ,\\
	R_c=&\frac{\Gamma(K^{-}p\to\text{charged particles})}
	{\Gamma(K^{-}p\to\text{all})}\sim 0.664\pm 0.011\ , \\
	R_n=&\frac{\Gamma(K^{-}p\to\pi^{0}\Lambda)}
	{\Gamma(K^{-}p\to\text{neutral particles})}\sim 0.189\pm 0.015 \ .
    \end{split}
    \nonumber
\end{equation}
Without the symmetry breaking terms,
we find the optimal value $a=-1.96$ (A).  
Now including the symmetry breaking term, the optimal value 
takes $a = -1.59$ (B).  
The calculated threshold values are presented in Table~\ref{tbl:branch}.  
From the table, we see that
the agreement with data is improved by including 
the symmetry breaking effect.
Note that this improvement is achieved
without new free parameters.

\begin{table}[tbp]
    \centering
\begin{ruledtabular}
    \begin{tabular}{cccc}
         & $\gamma$ & $R_c$ & $R_n$  \\
        \hline
        experiment  & $2.36\pm0.04$ & $0.664\pm 0.011$
	& $0.189\pm 0.015$  \\
        \hline
        (A) & 1.80  & 0.624 & 0.225  \\
        (B) & 2.19  & 0.623 & 0.179  \\
	(C) & 2.35  & 0.626 & 0.172  \\
    \end{tabular}
\end{ruledtabular}
    \caption{Threshold branching ratios calculated
    by using $a=-1.96$
    without the SU(3) breaking interaction (A),
    $a=-1.59$ with the SU(3) breaking interaction (B),
    $a=-1.68$ with the SU(3) breaking interaction 
    and the physical $f$ (C).
    Experimental values are take from
    Refs.~\cite{Nowak:1978au,Tovee:1971ga}.}
    \label{tbl:branch}
\end{table}%

\begin{figure}
    \centering
    \includegraphics[width=8.3cm,clip]{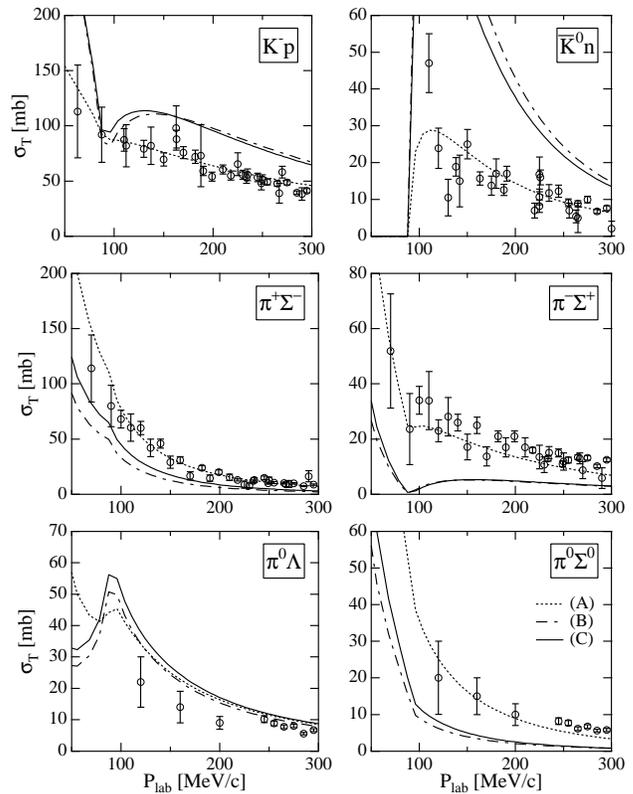}
    \caption{Total cross sections of 
    $K^{-}p$ scatterings ($S=-1$)
    as functions of  $P_{\text{lab}}$,
    the three-momentum of initial $K^{-}$
    in the laboratory frame.
    Dotted lines show 
    the results with $a=-1.96$ without SU(3) breaking (A),
    dash-dotted lines show the results including the SU(3) breaking 
    with $a=-1.59$ (B), and
    solid lines show the results including the SU(3) breaking 
    and the physical $f$ with $a=-1.68$ (C).
    Open circles with error bars are experimental data
    taken from Refs.~\cite{Mast:1976pv,
    Ciborowski:1982et,Bangerter:1981px,
    Mast:1975sx,Sakitt:1965kh,
    PL16.89,PRL14.29,
    PL21.349,NC16.848}.}
    \label{fig:S-1breakcross}\vspace{0.5cm}
\end{figure}

Using these optimal values, we calculate the cross sections of 
$K^- p \to$ (various channels) and plot them in Fig.~\ref{fig:S-1breakcross}.  
Results of (A) are shown by dotted lines and those of 
(B) by dash-dotted lines.
For (A), the agreement with data is still good, which is 
the well known result of the chiral unitary model~\cite{ 
Kaiser:1995cy,Oset:1998it,Oller:2000fj}.  
Originally the subtraction constants in the $S=-1$ channel
are not very much dependent on the channels and 
take values around $a_{i}\sim -2$ as shown in Table \ref{tbl:subtractions}.
Now including the symmetry breaking terms (B), 
we find that agreement with data becomes worse
(dash-dotted lines), contrary to our expectation, 
although the threshold branching ratios are better reproduced.  

\begin{figure}
    \centering
    \includegraphics[width=6cm,clip]{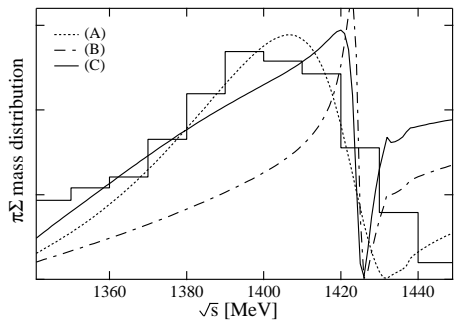}
    \caption{Mass distributions of the $\pi \Sigma$ channel
    with $I=0$.
    Dotted line shows 
    the result with $a=-1.96$ (A),
    dash-dotted line shows the result including the SU(3) breaking 
    with $a=-1.59$ (B), and
    solid line shows the result including the SU(3) breaking 
    and the physical $f$ with $a=-1.68$ (C).
    Histogram are experimental data
    taken from Ref.~\cite{Hemingway:1985pz}.}
    \label{fig:S-1breakmdist}
\end{figure}

In Fig.~\ref{fig:S-1breakmdist}
we show the $\pi \Sigma$ mass distribution,
in order to investigate the $\Lambda(1405)$ resonance.
For (A) we obtained the dotted curve which agrees well 
with experimental data.  
If we include the symmetry breaking terms (B), once again, 
the agreement becomes worse as shown by dash-dotted line.  
A sharp peak is pronounced around $\sqrt{s} = 1420$ MeV, 
in obvious contradiction with the observed spectrum.  

We also perform calculations with the inclusion of
another source of the SU(3) flavor breaking,
that is, the meson decay constants. 
We use the empirical values of the decay constants:
$f_{\pi}=93\text{ MeV},\;
f_{K}=1.22f_{\pi},\;f_{\eta}=1.3f_{\pi}$.
The optimal value of the subtraction constant $a$ in this case
is $a = -1.68$ (C) to reproduce the threshold ratios.
The results are shown in Figs.\ref{fig:S-1breakcross},
\ref{fig:S-1breakmdist} with
the solid lines. While the inclusion of the SU(3) breaking on 
the meson decay constants does not make drastic improvement 
in the total cross sections of the $K^{-}p$ scatterings
as shown in Fig.\ref{fig:S-1breakcross},
the shape of the peak in the $\pi\Sigma$ mass distribution becomes milder.
However, the improvement is not enough to reproduce the experimental spectra.

\begin{figure}
   \centering
   \includegraphics[width=8.3cm,clip]{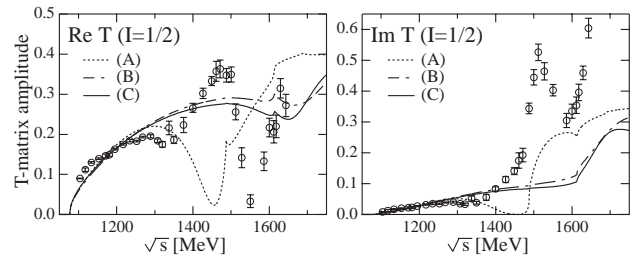}
   \caption{Real and imaginary parts of the $S_{11}$ T-matrix amplitudes of
   $\pi N\to \pi N$.
   Dotted lines show
     the results with $a=0.53$ (A),
   dash-dotted lines show the results
     including the SU(3) breaking interaction
     with $a=1.33$ (B), and
   solid lines show the results including the SU(3) breaking 
   and the physical $f$ with  $a=2.24$ (C).
   Open circles with error bars are experimental data
   taken from Refs.~\cite{CNS}.}
   \label{fig:S0breaktmat}
\end{figure}

We perform similar analyses for the $\pi N$ scattering for the
$S=0$ channel. 
At first we use the common subtraction constant $a=-1.96$ obtained
in the $S=-1$ channel without the symmetry breaking,
since it reproduces the $\Lambda(1405)$ property well and we 
want to check the SU(3) flavor symmetry. 
Then the attractive force between
the mesons and baryons is so strong that an unexpected resonance has been
generated at around $\sqrt s \simeq 1250$ MeV.
Therefore we choose the values of $a$ for $S=0$ by fitting the $S_{11}$
scattering amplitudes of the $\pi N$ channel up to the energy 
$\sqrt{s} \sim 1400$ MeV.
We show the calculated scattering amplitudes of the
$S_{11}$ $\pi N$ channel in Fig.~\ref{fig:S0breaktmat}
for the following three cases:
(A) $a = 0.53$ without SU(3) breaking,
(B) $a = 1.33$ with SU(3) symmetry breaking
and
(C) $a = 2.24$ with physical meson decay constants.
In all cases, the scattering amplitudes and cross sections 
(we do not show the cross sections here) are not well reproduced.
The results of (A) seems to have some structure around the 
$N(1535)$ energies, but it is far from the observed amplitude.
Reasonable agreement with data is
achieved only when channel dependent subtraction constants are
introduced as shown in Table~\ref{tbl:subtractions}~\cite{Inoue:2001ip}.

In this work, motivated by
the channel dependence of the parameters
and symmetry consideration,
we have tried to reproduce
the observed cross sections 
and the resonance properties
using a single subtraction constant.
In the $S=-1$ channel, without the symmetry breaking terms,
$a = -1.96$ is determined by the threshold 
branching ratios of the $K^- p$ scatterings.
With this parameter (A), 
the total cross sections of the $K^- p$ scatterings are 
reproduced well, as well as the mass distribution for $\Lambda(1405)$ is.
(See Figs.\ref{fig:S-1breakcross} and \ref{fig:S-1breakmdist}.)
This value is closed to the $a\sim -2$ corresponds to
$\Lambda = 630$ MeV in the three-momentum cut-off
reguralization~\cite{Oller:2000fj}.
The elementary interaction of the $\bar K N$ system 
is sufficiently attractive,
and a resummation of the coupled channel interactions 
generates the $\Lambda(1405)$ resonance
at the correct position,
by imposing the unitarity condition
with the natural value for the cut-off parameter.
Hence the wave function of $\Lambda(1405)$ is largely
dominated by the $\bar K N$ component.

On the other hand, in the $S=0$ channel,
if one uses the natural value for the subtraction constant 
as in the $S=-1$ channel,
the attraction of the meson-baryon interaction becomes so 
strong that an unexpected resonance is generated at around
$\sqrt s  \simeq 1250$ MeV. 
Therefore, repulsive component is
necessary to reproduce the observed $\pi N$ scattering.
However, with the fitted value $a\sim 0.5$,
the $N(1535)$ resonance
is not generated.

From the above observation, we see that
the unitarized amplitudes are very sensitive
to the attractive component of the interaction.
Even including the SU(3) breaking terms,
the interaction derived from the chiral Lagrangian alone
do not describe all scattering amplitudes simultaneously.
Both the fundamental interaction and the subtraction constants
are important in order to reproduce proper results.
For smaller $a$, the interaction becomes more attractive,
and for larger $a$, less attractive.
For $S=0$, we need to choose $a\sim 0.5$
in order to suppress the attraction
from the $\pi N$ interaction in contrast with
the natural value $a\sim -2$ in the $S=-1$ channel.
Therefore, it is not possible to reproduce both the
$\Lambda(1405)$ resonance properties and the low energy $\pi N$ 
scattering with a common subtraction constant.

At this point, it is useful to discuss slightly in detail
the structure of the $\Lambda(1405)$ resonance.
Although the properties of $\Lambda(1405)$ has not been
reproduced well with the SU(3) breaking terms as shown in the
$\pi\Sigma$ mass distribution (Fig.\ref{fig:S-1breakmdist}),
we still have found the two poles for $\Lambda(1405)$ in the scattering 
amplitudes in the second Riemann sheet.
The property of the two poles are investigated 
recently in detail and is related to the SU(3) structure of 
the meson and baryon
states~\cite{Oller:2000fj,Jido:2002yz,Garcia-Recio:2002td,Jido:2003cb}.
In the present study, we find
$z_1(B)=1424+1.6i$ and $z_2(B)=1389+135i$
for the parameter (B).
The pole $z_1$, which is located very close to the real axis,
is responsible for the sharp peak.  
When the SU(3) breaking
of the meson decay constants is introduced (C),
the poles are $z_1(C)=1424+2.6i$ and $z_2(C)=1363+87i$,
where $z_1$ is still close to the real axis,
while $z_2$ moves significantly.

The shape of the $\pi\Sigma$ mass distribution is strongly
influenced by the location of the poles.
In this case, the poles $z_2$ 
is sensitive to the pion decay constant.
Since the resonance of $z_2(B)$
has a strong coupling to the $\pi \Sigma$ channel~\cite{Jido:2003cb},
the resonance properties are very much affected by the
$\pi \Sigma$ interaction.
In the chiral Lagrangian, the interaction is attractive as in the
Weinberg-Tomozawa term, which contains a coupling strength proportional
to the inverse square of the pion decay constant.
Therefore, by changing the decay constant from the SU(3) 
averaged value (107 MeV, case B)
to the physical value (93 MeV, case C), the strength of the attractive
$\pi \Sigma$ interaction is enhanced by $\sim 30$ \%.
This shifts the real part of $z_2$ to the lower side.
At the same time this reduces the phase space and hence the
imaginary part decreases.

To summarize shortly, we have studied the flavor SU(3) symmetry 
breaking effect in the meson-baryon scatterings in the chiral unitary 
model. A reasonable prescription from symmetry consideration by
including the symmetry breaking mass terms, which appear in the 
next-to-leading order of the chiral expansion,
make theoretical predictions worse.
So far, except for the use of channel dependent subtraction constants, 
we do not know what would resolve this problem. 
In the present framework,
the role of the subtraction constants is very important.

A better understanding may be provided by
introducing genuine resonance components.   
Very naively such states could be quark originated as expected 
from the success of the quark model for baryon resonances.  
Full coupled channel studies of meson-baryon and quark degrees of 
freedom would be useful in order to resolve the problem 
discussed in the present study.  
Such an analysis will provide more microscopic understanding 
for the resonance structure.

\begin{acknowledgments}
We would like to thank 
Profs.\ E.\ Oset, H.\ -Ch.\ Kim and W.\ Weise for useful
discussions.
\end{acknowledgments}

\end{document}